# Scale-free correlations in bird flocks

Andrea Cavagna[1,2]★, Alessio Cimarelli[1,3], Irene Giardina[1,2]★, Giorgio Parisi[1,3,4]★, Raffaele Santagati[1,3], Fabio Stefanini[1,3◇], Massimiliano Viale[3]

[1] Centre for Statistical Mechanics and Complexity (SMC), CNR – INFM, Dipartimento di Fisica, Universita' di Roma 'La Sapienza', Piazzale Aldo Moro 2, 00185 Roma, Italy

[2] Istituto Sistemi Complessi (ISC), CNR, via dei Taurini 19, 00185 Roma, Italy

[3] Dipartimento di Fisica, Universita' di Roma 'La Sapienza', Piazzale Aldo Moro 2, 00185 Roma, Italy

[4] Sezione INFN, Universita' di Roma 'La Sapienza', Piazzale Aldo Moro 2, 00185 Roma, Italy

★ Corresponding authors:

andrea.cavagna@roma1.infn.it

irene.giardina@roma1.infn.it

giorgio.parisi@roma1.infn.it

◇ Current address:

Institut für Neuroinformatik, Universität Zürich, Winterthurerstrasse 190, CH-8057 Zurich, Switzerland

**From bird flocks to fish schools, animal groups often seem to react to environmental perturbations as if of one mind. Most studies in collective animal behaviour have aimed to understand how a globally ordered state may emerge from simple behavioural rules. Less effort has been devoted to understanding the origin of collective response, namely the way the group as a whole reacts to its environment. Yet collective response is the adaptive key to survivor, especially when strong predatory pressure is present. Here we argue that collective response in animal groups is achieved through scale-free behavioural correlations. By reconstructing the three-dimensional position and velocity of individual birds in large flocks of starlings, we measured to what extent the velocity fluctuations of different birds are correlated to each other. We found that the range of such spatial correlation does not have a constant value, but it scales with the linear size of the flock. This result indicates that behavioural correlations are scale-free: the change in the behavioural state of one animal affects and is affected by that of all other animals in the group, no matter how large the group is. Scale-free correlations extend maximally the effective perception range of the individuals, thus compensating for the short-range nature of the direct inter-individual interaction and enhancing global response to perturbations. Our results suggest that flocks behave as critical systems, poised to respond maximally to environmental perturbations.**

**Introduction**

Of all distinctive traits of collective animal behaviour the most conspicuous is the emergence of global order, namely the fact that all individuals within the group synchronize to some extent their behavioural state [1,2,3]. In many cases global ordering amounts to an alignment of the individual directions of motion, as in bird flocks, fish schools, mammal herds and in some insect swarms [4,5,6]. Yet, global ordering can affect also other behavioural states, as it happens with the synchronous flashing of tropical fireflies [7] or the synchronous clapping in human crowds [8].

The presence of order within an animal group is easy to detect. However, order may have radically different origins, and discovering what is the coordination mechanism at the basis of order is not straightforward. Order can be the effect of a top-down centralized control mechanisms (for example, due to the presence of one or more leaders), or it can be a bottom-up self-organized feature emerging from local behavioural rules [9]. Distinguishing between these two types of global ordering is not trivial. In fact, the prominent difference between the centralized and the self-organized paradigm is not order, but response.

Collective response is the way a group as a whole reacts to its environment. For gregarious animals under strong predatory pressure collective response is vital. The remarkable thing about a flock of birds is not merely the globally ordered motion of the group, but the way the flock dodges a falcon's attack. Collective response is the trademark of self-organized order as opposed to centralized one. Consider a group where all individuals follow a leader. Such system is strongly ordered, as everyone moves in the same direction. Yet, there is no passing of information from individual to individual and hence behavioural fluctuations are independent: the change of direction of one animal (different from the leader) has very little influence on that of other animals, due to the centralized nature of information transfer. As a consequence, collective response is very poor: unless detected directly by the leader, an external perturbation does not elicit a global reaction by the group. Response, unlike order, is the real signature of self-organization. Centralized global order, on the contrary, does not grant collective response.

In self-organized groups the efficiency of collective response depends on the way individual behavioural changes, typically forced by localized environmental perturbations, succeed in modifying the behaviour of the whole group. This key process is ruled by behavioural correlations. Correlation is the expression of an indirect information transfer mediated by the direct interaction between the individuals: two animals that are outside their range of direct interaction (be it visual, acoustic, hydrodynamic or any other) may still be correlated if information is transferred from one to another through the intermediate interacting animals. The turn of one bird attacked by a predator has an influence not only over the neighbours directly interacting with it, but also over all birds that are correlated to it. Correlation measures how the behavioural changes of one animal influences that of other animals across the group. Behavioural correlations are therefore ultimately responsible for the group's ability to respond collectively to its environment.

Of course, behavioural correlations are the product of inter-individual interaction. Yet interaction and correlation are different things and they may have a different spatial span. Interaction is local in space and its range is typically quite short. A former study [10] showed that in bird flocks the interaction range is of the order of few individuals (about seven). On the other hand, the correlation length, namely the spatial span of the correlation, can be significantly larger than the interaction range, depending chiefly on the level of noise in the system. An elementary example is the game of telephone: a player whispers a phrase into her neighbour's ear. The neighbour passes on the message to the next player and so on. The direct interaction range is equal to one, whereas the correlation length, i.e. the number of individuals the phrase can travel before being corrupted, can be significantly larger than one, depending on how clearly the information is transmitted at each step.

Although the correlation length is typically larger than the interaction range, in most biological and physical cases it is significantly smaller than the size of the system. For example, in bacteria the correlation length was found to be much smaller than the size of the swarm [11,12]. In this case parts of the group that are separated by a distance larger than the correlation length are by definition independent from each other and therefore react independently to environmental perturbations. Hence, the finite scale of the correlation necessarily limits the collective response of the group.

However, in some cases the correlation length may be as large as the entire group, no matter the group's size. When this happens we are in presence of scale-free correlations [13,14]. The group cannot be divided into independent sub-parts, because the behavioural change of one individual influences and is influenced by the behavioural change of all other individuals in the group. Scale-free correlations imply that the group is, in a strict sense, different from and more than the sum of its parts [15]. The effective perception range of each individual is as large as the entire group and it becomes possible to transfer undamped information to all animals, no matter their distance, making the group respond as one. Here, we provide experimental evidence that bird flocks exhibit scale-free correlations and we discuss under what conditions such correlations may arise in animal groups.

## Results

We measured the three-dimensional positions and velocities of individual birds within large flocks of starlings (*Sturnus vulgaris*) in the field [16]. Data were taken at sunset over a major roosting site in Rome in the winter months of 2005-2007. Analyzed flocks ranged from 122 to 4268 individuals [17,18,19], two orders of magnitude larger than any previously studied animal group in three dimensions. The degree of global ordering in a flock is measured by the so-called polarization $\Phi$,

$$\Phi = \left\| \frac{1}{N} \sum_{i=1}^{N} \frac{\vec{v}_i}{\|\vec{v}_i\|} \right\|, \qquad (1)$$

where $\vec{v}_i$ is the velocity of bird $i$ and $N$ is the total number of birds within the flock. The polarization is zero if the individual velocities are pointing in different directions, while it is close to one if most of them are nearly parallel. In fact, a nonzero value of $\Phi$ means that there is net motion of the centre of mass. Polarization is therefore used as a standard measure of global order in the study of collective animal behaviour [20,21]. In all analyzed flocks we found very high values of the polarization (see Table S1 in the Supporting Information). The average value over all 24 flocks is $\Phi=0.96\pm0.03$ (standard deviation), confirming the visual impression of strongly ordered birds' velocities (see Fig.1A for a two-dimensional projection of the individual three-dimensional velocities).

However, as we have stressed above, order tells us little about collective response. To learn something about response we must study how the fluctuations in the behavioural state (in this case the velocity) of one bird are correlated to those of another bird. Let us introduce for each bird $i$ the fluctuation around the mean flock's velocity,

$$\vec{u}_i = \vec{v}_i - \frac{1}{N} \sum_{k=1}^{N} \vec{v}_k, \qquad (2)$$

which is nothing else than the bird's velocity in the centre of mass reference frame. The spatial mean of the velocity fluctuations is zero by construction,

$$\sum_{i=1}^{N} \vec{u}_i = 0. \qquad (3)$$

Relation (3) encodes the obvious fact that there cannot be overall net motion in the centre of mass reference frame.

In Fig.1C we report the probability distribution of the modulus of the full velocity (the speed) and of the modulus of the velocity fluctuations in a typical flock. The modulus of the fluctuations $\vec{u}_i$ is on average much smaller than that of the velocities $\vec{v}_i$. This is expected, because the polarization is very large and thus the fluctuations around the mean are small. Yet, despite their small values, the velocity fluctuations contain a great deal of information, as it is clear from an inspection of Fig.1B. Even in such two-dimensional projection of a three-dimensional flock it is possible to detect the presence of large domains where the fluctuations are nearly parallel to each other (see Fig.S1 in the Supporting Information for another flock). The existence of these domains is not a consequence of the fact that birds are all flying in the same direction, because the overall centre of mass velocity has been subtracted in equation (2). Hence, what Fig.1B shows is the presence of strong spatial correlations: the change of heading of a bird within one of these domains is highly correlated to that of all birds within the same domain. Previous studies on starling flocks have shown that each bird interacts on average with 7 neighbours [10]. From Fig.1B it is clear that the correlated domains contain much more than 7 birds. Hence, the span of spatial correlation is significantly larger than the interaction range. To quantify the size of the domains we define in three dimensions the correlation function of the fluctuations,

$$C(r) = \frac{\sum_{ij} \vec{u}_i \cdot \vec{u}_j \, \delta(r - r_{ij})}{\sum_{ij} \delta(r - r_{ij})} \quad , \quad (4)$$

where $\delta(r - r_{ij})$ is a smoothed Dirac delta-function selecting pairs of birds at mutual distance $r$ (see Methods) and $\vec{u}_i \cdot \vec{u}_j = u_i^x u_j^x + u_i^y u_j^y + u_i^z u_j^z$. The correlation function measures the average inner product of the velocity fluctuations of birds at distance $r$. A large value of $C(r)$ implies that the fluctuations are nearly parallel, and thus strongly correlated. Conversely, when the fluctuations are anti-parallel, and therefore anti-correlated, the correlation function has a negative value. On the other hand, when the fluctuations are uncorrelated, pointing in random directions, the correlation function averages to zero.

The typical form of $C(r)$ in starling flocks is reported in Fig.2A (for other samples see Fig.S2 in the Supporting Information). At short distances the correlation is close to 1, it decays with increasing $r$, becoming negative at large inter-individual distances (for $r$ larger than the flock's size $C(r)$ is no longer defined). Such behaviour indicates that within a flock there is either strong correlation (short distance) or strong anti-correlation (large distance), whereas in no range of $r$ the correlation function is consistently equal to zero, as one would expect in the case of absence of correlation. This phenomenon can be seen at a qualitative level in Fig.1B: each correlated domain has an anti-correlated domain with opposite fluctuations. Their mutual negative correlation must not be misunderstood for an absence of correlation: the latter case would imply a random distribution of orientations and therefore a correlation function equal to zero over a finite interval, at variance with the correlation functions we find. We note that the presence of correlated/anti-correlated domains pairs, and therefore the fact that the correlation function is positive and negative, is a trivial consequence of the fact that the spatial average of $\vec{u}_i$ is zero (equation (3)). However, what is highly not trivial is the fact that just two domains (the minimum number) span the entire system.

In order to explain the behaviour of $C(r)$ we introduce the correlation length $\xi$, which can be defined as the zero of the correlation function,

$$C(r = \xi) = 0 \quad . \quad (5)$$

The value of $\xi$ coincides with the average size of the correlated domains (see Methods and Fig.S3 in the Supporting Information). Indeed, the fact that the correlation function changes sign at $r = \xi$ translates the fact that when we increase $r$ we pass from correlated to anti-correlated domains in the flock. What is the typical value of $\xi$? A former study showed that the interaction range has a constant value in units of birds (about 7 individuals), rather than in units of meters [10]. Hence, one may naively expect that the correlation length too has a constant 'topological' value (units of individuals), rather than a constant metric value. What we find is however completely different, and somewhat surprising: we measured the correlation length in all analysed flocks and found that $\xi$ does not have a constant value, neither in unit of birds, nor in unit of meters. Rather, the correlation length grows linearly with the size of the flock $L$ (Fig.2C). Accordingly, correlated domains in starling flocks are larger the larger the flock.

A correlation length that is proportional to the system size implies that correlations are scale-free. Let us briefly recall how this works. In general, we can write the leading contribution to the correlation function in the following way [13, 14],

$$C(r) = \frac{1}{\xi^\gamma} g(r/\xi) \quad , \quad (6)$$

where $g(x)$ is a dimensionless scaling function. As we have seen, we find that the correlation length grows with the flock's size $L$; this result can be formalized as,

$$\xi(bL) = b\, \xi(L) \quad , \quad (7)$$

where $b$ is a generic scaling factor. By plugging (7) into the general relation (6) we obtain,

$$C(r;L) = b^\gamma\, C(br;bL) \quad . \quad (8)$$

By choosing $b=1/r$, we finally obtain the following form for the correlation function in starling flocks,

$$C(r;L) = \frac{1}{r^\gamma} f(r/L) \quad . \quad (9)$$

Equation (9) explains the meaning of the expression 'scale-free': the correlation between birds does not have any characteristic length scale apart from the trivial one fixed by the size of the flock, $L$. The correlation length $\xi$ defined above is not an intrinsic length scale, for it is proportional to $L$. The scaling function $f(r/L)$ in (9) embodies the effect of the flock's finite size on the correlation function. To get rid of such effect and find the *asymptotic* correlation function $C_\infty(r)$, we simply ask what is the correlation between two birds at distance $r$ within a very large flock; to answer this question we perform the limit $L \to \infty$ in (9) and get,

$$C_\infty(r) = \frac{1}{r^\gamma} f(0) \approx \frac{1}{r^\gamma} \quad . \quad (10)$$

Equations (9) and (10) make the main point of our work: the empirical observation that the correlation length is proportional to $L$ (Fig.2C and equation (7)) implies that correlations in starling flocks are scale-free and that the asymptotic correlation function is a power law.

What is the value of $\gamma$? The sharpest way to work out the value of this exponent is to calculate the derivative of the finite size correlation function with respect to the rescaled variable $x = r/\xi$. According to equation (6), when we evaluate this derivative at the zero of the correlation function, i.e. at $x =1$, we obtain,

$$C'(x=1) = \frac{1}{\xi^\gamma} g'(1) \approx -\frac{1}{\xi^\gamma} \approx -\frac{1}{L^\gamma} \quad . \quad (11)$$

Hence, the rescaled correlation function at its zero should flatten (lower derivative) in larger flocks. In Fig.3A we plot several correlation functions *vs.* the rescaled variable $x = r/\xi$: up to experimental error the curves seem to collapse quite well one onto the other, with no clear evidence of a flattening of the derivative for larger flocks, indicating that $\gamma$ in eq. (11) has a very small value. In the inset of Fig.3C we report for all the analyzed flocks the absolute value of the derivative in zero *vs*. ξ. What we observe is indeed a *very* weak decrease of the derivative with increasing correlation length. The best fit to equation (11) gives a very small exponent ($\gamma = 0.19 \pm 0.08$, with Reduced Chi Square-RCS= 0.045), but the data are equally compatible with a logarithmic decay (RCS=0.040), and even with a constant value of the derivative, equivalent to γ=0 or no decay (RCS=0.059). The data barely span one order of magnitude, so it would be unwise to commit to any of these fits. However, what the data positively demonstrate is that the value of γ is very low indeed.

This result is rather startling. In a non-scale-free system, the asymptotic correlation between two individuals drops to zero when their distance gets larger than ξ. On the contrary, in a scale-free system the asymptotic correlation is never zero, but it nevertheless decays, albeit as a power law, $1/r^\gamma$. However, if gamma is barely different from zero, as it seems to be the case in starling flocks, then the asymptotic correlation (i.e. the correlation within infinitely large flocks) practically *does not decay with the distance*. From equation (8) we see that an almost zero value of γ implies that two birds 1m apart in a 10m wide flock are as strongly correlated as two birds 10m apart in a 100m wide flock. Behavioural correlations in starling flocks are therefore not simply scale-free, but in fact they are unusually long-ranged.

To better understand the significance of scale-free correlations it is useful to see what happens in the non-scale-free case. To this aim we use synthetic data (see Methods). In each flock we substitute the actual velocity fluctuations with a set of synthetic random vectors correlated according to the following asymptotic correlation function,

$$\hat{C}_\infty(r) = \frac{1}{r^\gamma} \exp(-r/\lambda) \quad . \quad (12)$$

We have used the hat to distinguish this synthetic correlation function from the biological one. In contrast with (10), the synthetic correlation function (12) is clearly not scale-free, as the decay rate λ (which we can arbitrary tune) fixes a spatial scale, and the correlation is exponentially suppressed for $r > \lambda$. Hence, the finite size correlation function $\hat{C}(r;L)$, calculated according to definition (4), does not obey the scale-free relation (9). When λ is small, domains are also small (Fig.4A) and $\hat{C}(r;L)$ is consistently equal to zero beyond distances of order λ (Fig. 4C). This means that portions of the flock separated by a distance larger than λ are uncorrelated and behave independently. The correlation length ξ is a constant, approximately equal to λ, and it does not scale with *L*. As we increase λ the size of the synthetic domains grows and the correlation function becomes more and more long-ranged (Fig.4C), but nothing qualitative changes as long as λ<*L*.

On the other hand, if the decay rate λ is larger than the size *L* of the flock, then all possible values of the inter-individual distance *r* are much smaller than λ, and therefore the exponential in (12) is always well approximated by 1. In this case the asymptotic correlation function of the synthetic data decays as a scale-free power law,

$$\hat{C}_\infty(r) \approx \frac{1}{r^\gamma} \quad , \quad (13)$$

exactly as in the case of real flocks, eq.(10). We therefore expect that in the scale-free limit (13) the synthetic finite-size correlation function must become equal to that of real flocks, provided that we choose a value of γ that is small enough. This is exactly what we find: the synthetic correlation function (Fig. 4C) and the synthetic domain size and correlation length (Figs. 4B and 4D) become barely distinguishable from their biological counterparts when the scale-free form (15) holds, i.e. in the regime $\lambda > L$.

So far we have studied the correlations of the *orientation* of the velocity (equation (4)). However, when we compute the correlation function of the *speed* (i.e. the modulus of the velocity - see Methods for details) we find an identical linear scaling with $L$ of the corresponding correlation length (Figs. 2B and 2D). Hence, speed correlations are scale-free, exactly as orientation correlations. Moreover, the analysis of $\gamma$ (Fig. 3B and inset) gives a very small value for this exponent, exactly as for the orientation ($\gamma = 0.19 \pm 0.11$ - RCS=0.10; logarithmic decay, RCS=0.068; constant-no decay, RCS=0.097). Therefore, speed fluctuations too are very long-ranged, almost not decaying with the distance.

The speed is a stiffer mode than the orientation, as it is more costly for a bird to change its speed (accelerate/decelerate) than its heading. Hence, the fact that both orientation and speed are scale-free correlated means that birds are able to transfer across the flock their whole dynamical state. In flocking, any external perturbation, and in particular predation, is likely to directly cause a change of velocity (direction, modulus, or both) of a small subset of birds that first detect the perturbation. Such localized change must transmit to the whole flock to produce a collective response. We do not focus here on the time-scale for this to happen, but on the very possibility for the information to reach the whole group, irrespective of the time needed to do this. In a group with finite correlation length $\xi$ the fluctuation of the dynamical state gets damped beyond $\xi$. On the contrary, in a flock where correlations in both speed and orientation are scale-free, and where the power-law exponent $\gamma$ is very small, information can reach the whole group without damping. Therefore, scale-free correlations are the key to collective response in bird flocks.

**Discussion**

Significant spatial correlations have already been observed in bacteria swarms [11,12]. In [12] it was found that for large enough densities of the bacterial swarm the correlation length becomes several body-lengths long. However, in bacteria the correlation function decays exponentially and the correlation length remains much shorter than the swarm size: correlation, as well as interaction, is short-ranged. What we find in starling flocks is different: the correlation function is a scale-free power-law and the correlation length scales with the group's size; hence, interaction is short-ranged, but correlation is long-ranged. If a correlation length larger than the interaction range is likely to be a common trait of self-organized groups, scale-free correlations seem the landmark of a qualitatively different kind of collective animal behaviour, characterized by a superior level of collective response.

Under what conditions scale-free correlations appear? What scale-free correlations teach us about the inter-individual coordination mechanism? First, there is no need to postulate the existence of complicated coordination mechanisms to explain scale-free correlations: simple behavioural rules based on imitation, as those used in most numerical models [20,21,22], are compatible with scale-free correlations. Indeed there are several statistical models based on simple alignment rules that develop scale-free correlations under certain circumstances [13]. The key point is not the rule, but the noise. Given a reasonable behavioural rule (for example, align your velocity to that of your neighbours), correlation strongly depends on the level of noise in implementing such rule. In a thermal system noise is due to the temperature, whereas in animal groups it is introduced by the inevitable individual error in obeying to any behavioural rule. Ordinarily, in self-organized systems the lower is the noise, the longer the range of the correlation. In this context order and correlation have a common origin: they are both large when the noise level in the system is low. Hence, it may be expected that bird flocks, which as we have seen are highly ordered, also exhibit strong correlations. In this case order, correlation and response would all be a consequence of the capability of flocking birds to obey a certain set of behavioural rules allowing very little tolerance, irrespective of the level of environmental perturbation the flock may undergo.

However, the relationship between noise and correlation may be more complex than that just described. In some cases, correlation (and hence response) reaches a maximum at a specific level of the noise. If noise is lowered below such critical level, order continues to grow, while correlation of the fluctuations actually decreases. This is what happens when a critical point is present. A classic example is ferromagnetism: below the critical temperature the global magnetization grows, but the local fluctuations around the global magnetization become less correlated. In this case order and correlation are decoupled: increasing the degree of order in the system (by lowering the noise below the critical point) makes the behavioural state of the individuals more stable, but also less sensitive to neighbouring behavioural changes. Such higher behavioural inertia depresses, instead of enhancing, the correlation and the global response of the group. Too much noise, on the other hand, equally destroys correlation, so that the system must contain just the right amount of noise to produce maximum response. For this reason, only at the critical point correlations are scale-free. In most physical systems criticality is obtained by tuning some external parameter regulating the noise (as the temperature) to its critical value. In the case of flocks, however, the critical value of the noise, i.e. of the random deviation from the coordination rules, must be evolutionary hard-wired into birds' behaviour.

Discriminating between the two scenarios above (very low noise vs. criticality) is difficult. We know too little about the actual inter-individual coordination mechanisms to conclude anything for sure. If scale-free correlations of a 'soft' degree of freedom as the orientation may be expected also off a critical point, the fact that a 'stiff' mode as the speed is scale-free correlated seems however to indicate that some kind of criticality might in fact be present in starling flocks. Indeed, as we discuss in the Supporting Information, scale-free correlations of a stiff degree of freedom are difficult to obtain by simply decreasing the noise in the system. Too low a noise level in a hard-to-change behavioural mode, as speed is, can cause an excessive behavioural inertia, which in turn depresses correlation and global response. For this reason criticality is perhaps a more likely scenario for our results. Whatever is the origin of the scale-free behaviour, though, the very low value of the exponent $\gamma$ that we find, i.e. the fact that the correlation is almost *not* decaying with the distance, is by far and large the most surprising and exotic feature of bird flocks. How starlings achieve such a strong correlation remains a mystery to us.

Criticality is not uncommon in biological systems made up of many interacting components. In particular, it has an important role in neurosciences: both experiments [23] and mathematical models [24] indicate that assemblies of neurons develop long-range correlations and large response as the result of criticality. There is an intriguing similarity with what we find in bird flocks: in both cases being critical is a way for the system to be always ready to optimally respond to an external perturbation, be it a sensory stimulus as in the case of neural assemblies, or a predator attack as in the case of flocks. Commonalities between animal groups and neural systems have already been noted and discussed in the literature [25,26]. Our empirical results, together with further study on the role of criticality in animal groups, may contribute to move the fascinating 'collective mind' metaphor to a more quantitative level.

## Materials and Methods

**Empirical observations.** Data were taken from the roof of Palazzo Massimo – Museo Nazionale Romano, in the city centre of Rome, in front of one of the major roosting sites used by starlings during winter. Birds spend the day feeding in the countryside and come back to the roost in the evening, about one hour before sunset. Before settling on the trees for the night, starlings gather in flocks of various sizes and perform what is called 'aerial display', namely an apparently purposeless dance where flocks move and swirl in a remarkable way. By using stereometric digital photogrammetry and computer vision techniques we reconstructed the individual 3D positions and 3D velocities in 24 flocking events. A flocking event is a series of consecutive shots of a flock at a rate of 10 frames-per-seconds. Analyzed flocks had different number of birds (from 122 to 4268 individuals) and different linear sizes (from 9.1m to 85.7m meters). The details of the 3D reconstruction of the positions can be found in Ref. [18, 19]. An algorithm similar to that used to solve the problem of the stereometric matching was used to perform the dynamical matching (or tracking), i.e. to associate to each bird's coordinates at time $t$ its corresponding coordinates at time $t+dt$. In our case the time interval between two consecutive reconstructions was $dt$=0.1sec. From this dynamical pair one can work out the velocity as the ratio between displacement and time. The efficiency of our dynamical matching algorithm, defined as the ratio between the number of matched birds and the original number of birds, is 0.77 on average.

**Correlation function.** The correlation function $C(r)$ defined in (4) is calculated by averaging the inner (or scalar) product of the velocity fluctuations of all pairs of birds with mutual distance in the interval $(r, r+dr)$, where $dr$ sets the discrete scale of $C(r)$. In this sense it must be interpreted the smoothed Dirac-delta in equation (4). The correlation function is normalized in such a way to give $C(r=0)=1$. The integral over $r$ between 0 and $L$ (size of the flock) of the numerator of equation (4) is zero thanks to equation (3),

$$\int_0^L dr \sum_{ij} \vec{u}_i \cdot \vec{u}_j \, \delta(r-r_{ij}) = \sum_{ij} \vec{u}_i \cdot \vec{u}_j = \sum_i \vec{u}_i \cdot \sum_j \vec{u}_j = 0 \quad . \quad (14)$$

As a consequence, the numerator in (4) *must* have a zero in the interval [0:$L$], and therefore the same holds for the whole function $C(r)$. Thanks to this condition we can define the correlation length $\xi$ as in equation (5). The correlation function of the velocity modulus, the speed, is defined as,

$$C_{sp}(r) = \frac{\sum_{ij} \varphi_i \cdot \varphi_j \, \delta(r - r_{ij})}{\sum_{ij} \delta(r - r_{ij})} \quad , \qquad (15)$$

where the delta-function has the same meaning as explained above and where,

$$\varphi_i = \|\vec{v}_i\| - \frac{1}{N} \sum_{k=1}^{N} \|\vec{v}_k\| \quad , \qquad (16)$$

is the (scalar) fluctuation of the speed with respect to the global mean. The same arguments used for the velocity fluctuations hold also for the speed fluctuations, so that $C_{sp}(r)$ must have a zero, $\xi_{sp}$.

**Size of the domains.** The correlation length $\xi$ provides a good estimate of the size of the correlated domains. To check this point we computed the size of the domains in an alternative way, by diagonalizing the covariance matrix $C_{ij} = \vec{u}_i \cdot \vec{u}_j$. The $N$-dimensional eigenvector $w^{max}$ relative to the maximum eigenvalue of this matrix defines the direction of maximal mutual alignment of the fluctuations, i.e. the average orientation of the largest correlated domain. Defining this eigenvector is useful, because if bird $i$ belongs to the correlated domain, then the $i$ component of the eigenvector $w^{max}$ is significantly different from zero. This is the rigorous way to identify the birds belonging to a correlated domain. Once the domain is defined, we calculate the domain's size using the median of the mutual distances of the birds belonging to it. In Fig.S2 we report the domains' size thus calculated as a function of the correlation length $\xi$. The clear linear correlation, with angular coefficient very close to 1, shows that $\xi$ is indeed a good estimate of the domains size.

**Synthetic random velocities.** At each instant of time a flock is characterized by a set of 3D coordinates (the birds positions $\vec{x}_i$) and of 3D vectors (the fluctuations $\vec{u}_i$ around the mean velocity). Given a flock, we keep the actual 3D positions, but replace the 3D fluctuations with a set of random vectors $\vec{w}_i$ (synthetic fluctuations), drawn with a distribution whose covariance matrix is given by,

$$\langle \vec{w}(x) \cdot \vec{w}(x+r) \rangle = \frac{\exp(-r/\lambda)}{(a+r)^\gamma} \quad . \qquad (17)$$

The length $\lambda$ sets the decay rate of the synthetic correlation, whereas the factor $a$ simply makes the correlation non-singular in $r = 0$. When $\lambda \gg L$ the exponential is always unity, and the correlation becomes a power law with exponent $\gamma$. Not any power makes a power-law scale-free, though. In three dimensions the power law is actually scale-free only for $\gamma < 3$, whereas if $\gamma > 3$ the correlation length does not scale linearly with $L$ and the correlation is effectively short-ranged. As we have seen in the main text, in order to have a good agreement with the biological data we need to use a very small value of this exponent. Practically speaking, any value $\gamma < 1$ gives synthetic results compatible with the biological ones, within the experimental error.

**Acknowledgements:** We acknowledge the help of N. Cabibbo and A. Orlandi in the early stages of this work, and of the whole STARFLAG-CNR experimental team during the first data-taking season. We are also indebted to W. Bialek, R. Bon, E. Branchini, C. Castellano, G. Cavagna, M. Cencini, I.D. Couzin, A. Gabrielli, D. Grunbaum, P.S. Krishnaprasad, J. Lorenzana, M. Magnasco, and G. Theraulaz for several helpful discussions. AC and IG are particularly grateful to T.S. Grigera for some key remarks on information transfer and to J. Gautrais for a critical reading of the manuscript. We thank R. Paris and M. Petrecca for granting the access to Palazzo Massimo, Museo Nazionale Romano, which was indispensable for the data taking. This work was partially financed by a grant from the European Commission under the STARFLAG project.

# REFERENCES

1. Parrish, J. K. & Hammer, W. M. (ed.) *Animal groups in three dimensions* (Cambridge Univ. Press, Cambridge, 1997).

2. Krause, J. & Ruxton, G.D. *Living in Groups* (Oxford University Press, Oxford, 2002).

3. Couzin, I. D. & Krause, J. Self-organization and collective behaviour in vertebrates. *Adv. Study Behav.* **32**, 1-75 (2003).

4. Okubo, A. Dynamical aspects of animal grouping: swarms, schools, flocks, and herds. *Adv. Biophys.* **22**, 1-94 (1986).

5. Emlen, J. T. Flocking behaviour in birds. *The Auk* **69**, 160-170 (1952).

6. Buhl, J. et al. From disorder to order in marching locusts. *Science* **312**, 1402-1406 (2006).

7. Buck, J. & Buck E. Mechanism of Rhythmic Synchronous Flashing of Fireflies. *Science* **159**, 1319-1327 (1968).

8. Néda Z., Ravasz E., Brechet Y., Vicsek T. & Barabási A.-L. Self-organizing processes: The sound of many hands clapping, *Nature* **403**, 849-850 (2000).

9. Camazine, S., Deneubourg, J.L., Franks, N.R., Sneyd, J., Theraulaz, G. & Bonabeau, E. *Self-organization in biological systems* (Princeton Studies in Complexity, Princeton University Press, Princeton, 2001)

10. Ballerini, M. et al, Interaction Ruling Animal Collective Behaviour Depends on Topological rather than Metric Distance: Evidence from a Field Study. *Proc. Natl. Acad. Sci. USA* **105**, 1232-1237 (2008).

11. Dombrowski, C., Cisneros L., Chatkaew S., Goldstein, R.E. & Kessler, J.O. Self-Concentration and Large-Scale Coherence in Bacterial Dynamics. *Phys. Rev. Lett*. **98**, 098103 (2004).

12. Sokolov, A., Aranson, I.S., Kessler, J.O. & Goldstein, R.E. Concentration dependence of the collective dynamics of swimming bacteria. *Phys. Rev. Lett*. **98**, 158102 (2007).

13. Wilson, K. Problems in physics with many scales of length, *Scientific American* **241**, 140-157 (1979).

14. Vicsek, T. *Fluctuations and Scaling in Biology* (Oxford University Press, Oxford, 2001).

15. Anderson, P.W. More is different, *Science* **177**, 393 – 396 (1972).

16. Feare, C. *The Starling* (Oxford Univ. Press, Oxford, 1984).

17. Ballerini, M. et al, An empirical study of large, naturally occurring starling flocks: a benchmark in collective animal behaviour. *Animal Behaviour* **76**, 201-215 (2008).

18. Cavagna, A. et al, The STARFLAG handbook on collective animal behaviour: Part I, empirical methods. *Animal Behaviour* **76**, 217-236 (2008).

19. Cavagna, A. et al, The STARFLAG handbook on collective animal behaviour: Part II, three-dimensional analysis. *Animal Behaviour* **76**, 237–248 (2008).

20. Vicsek, T., Czirok, A., Ben-Jacob, E. Cohen, I. & Shochet, O. Novel type of phase transition in a system of self-driven particles. *Phys. Rev. Lett*. **75**, 1226-1229 (1995).

21. Gregoire, G. & Chate, H. Onset of collective and cohesive motion. *Phys. Rev. Lett.* **92**, 025702 (2004).

22. Toner J., Tu Y. & Ramaswamy S. Hydrodynamics and phases of flocks. *Annals of Physics* **318**, 170-244 (2005).

23. Schneidman E., Berry M.J., Segev R. & Bialek W. Weak pairwise correlations imply strongly correlated network states in a neural population. *Nature* **440**, 1007-1012 (2006).

24. Magnasco M.O., Piro O. & Cecchi G.A. Self-Tuned Critical Anti-Hebbian Networks *Phys. Rev. Lett.* **102**, 258102 (2009).

25. Couzin, I. D. Collective minds. *Nature* **445**, 715 (2007).

26. Couzin, I. D. Collective cognition in animal groups. *Trends Cog. Sci*. **13**, 36-43 (2009).

# Figure 1

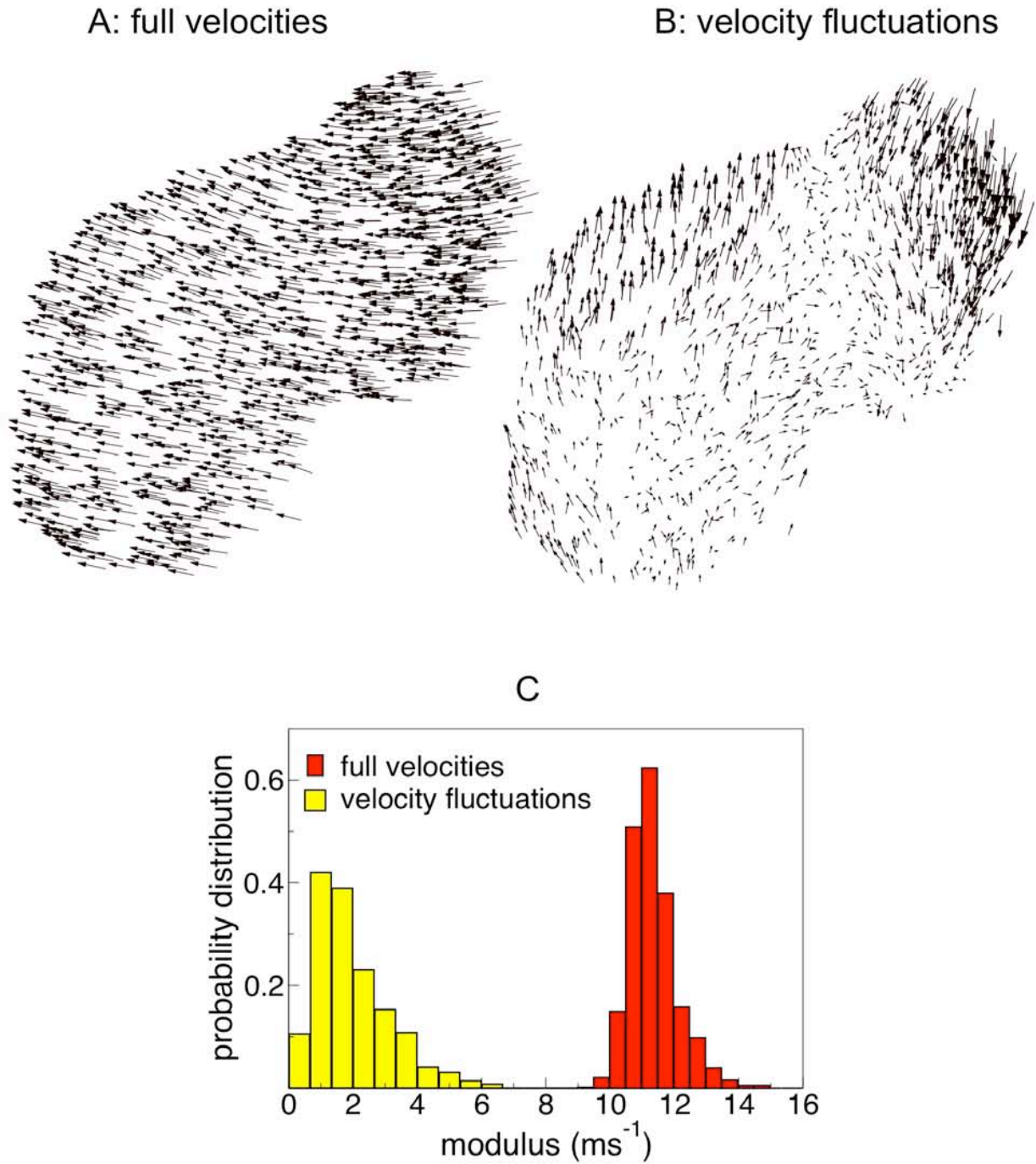

**Fig.1 A.** Two-dimensional projection of the velocities of the individual birds within a starling flock at a fixed instant of time (flock 28-10; 1246 birds, linear size $L$=36.5m). Vectors are scaled for clarity. The flock is strongly ordered and the velocities are all aligned. **B.** Two-dimensional projection of the individual velocity fluctuations in the same flock at the same instant of time (vectors scaled for clarity). The velocity fluctuation is equal to the individual velocity minus the centre of mass velocity, and therefore the spatial average of the fluctuations must be zero. Large domains of strongly correlated birds are clearly visible. **C.** Normalized probability distribution of the absolute value of the individual velocities and of the absolute value of the velocity fluctuations (same flock as in A and B). The velocity fluctuations are much smaller in modulus than the full velocities.

Figure 2

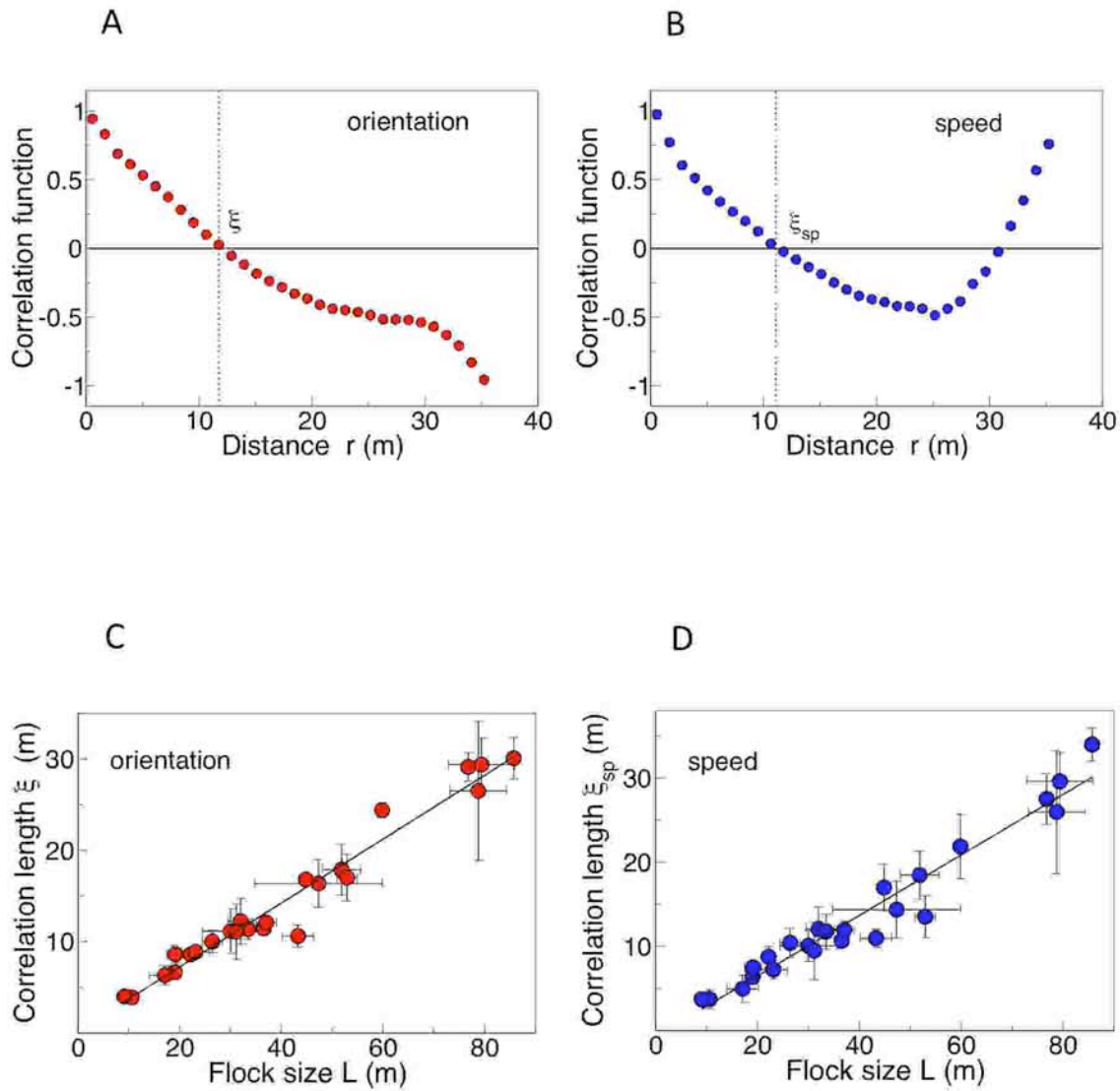

**Fig.2 A.** The correlation function $C(r)$ is the average inner product of the velocity fluctuations of pairs of birds at mutual distance $r$. This correlation function therefore measures to what extent the *orientations* of the velocity fluctuations are correlated. The function changes sign at $r=\xi$, which gives a good estimate of the average size of the correlated domains (flock 28-10). **B.** The correlation function $C_{sp}(r)$, on the other hand, measures the correlations of the fluctuations of the *modulus* of the velocity, i.e. the speed. This correlation function measures to what extent the variations with respect to the mean of the birds' speed are correlated to each other. The speed correlation function changes sign at a point $r=\xi_{sp}$, which gives the size of the speed-correlated domains (flock 28-10). Both correlation functions in panels A and B are normalized as to give $C(r=0)=1$. **C.** The orientation correlation length $\xi$ is plotted as a function of the linear size $L$ of the flocks. Each point corresponds to a specific flocking event and it is an average over several instants of times in that event. Error bars are standard deviations. The correlation length grows linearly with the size of the flock, $\xi = aL$, with a=0.35 (Pearson correlation test: n=24, r=0.98, $P<10^{-16}$), signalling the presence of scale-free correlations. **B**. Also in the case of the correlation function of the speed, the correlation length $\xi_{sp}$ grows linearly with the size of the flock, $\xi_{sp} = aL$, with a=0.36 (Pearson: n=24, r=0.97, $P<10^{-15}$). Error bars are standard deviations.

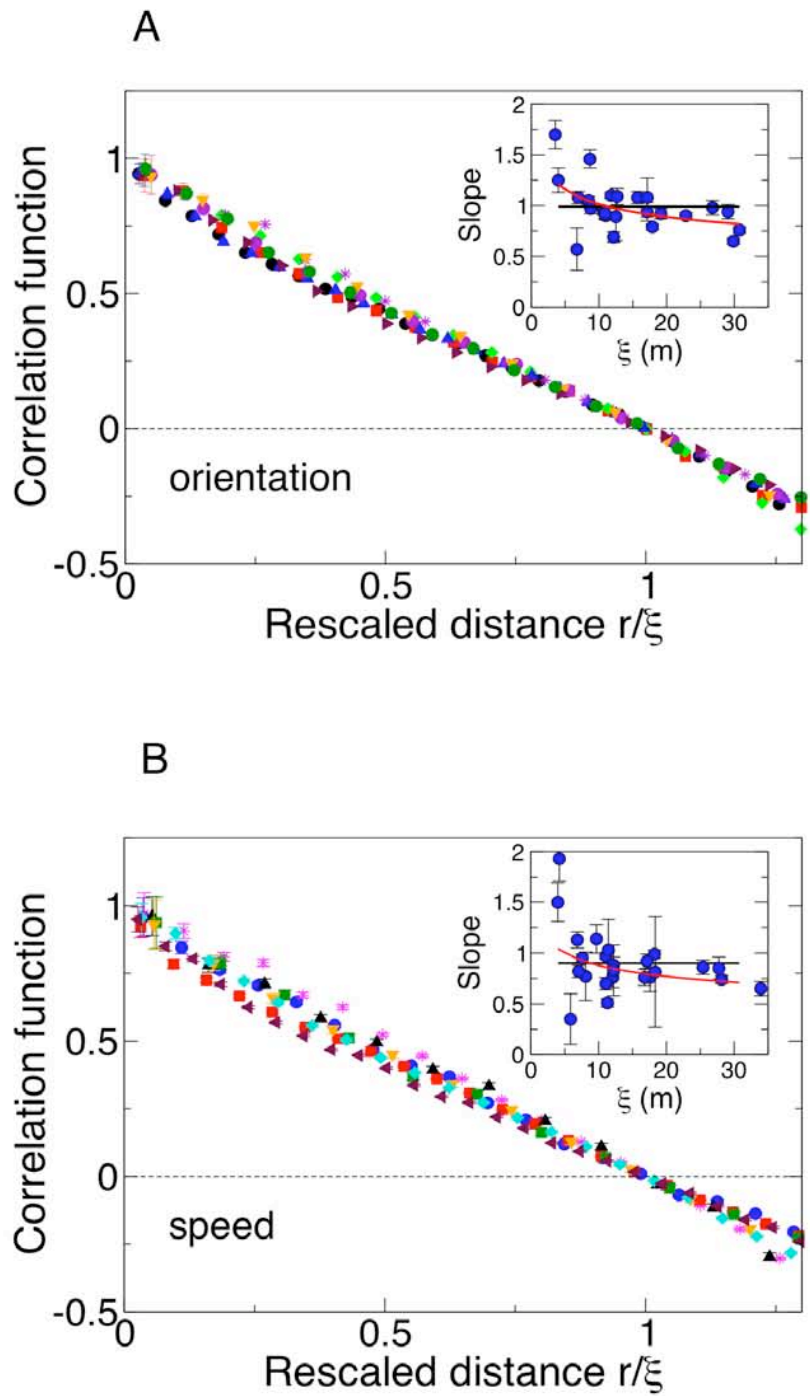

**Fig.3 A**. The correlation functions of all flocks are plotted *vs*. the rescaled variable $x = r/\xi$. Inset: the modulus of the derivative of the correlation function with respect to the rescaled variable $x$, evaluated at $x=1$, plotted *vs*. the correlation length $\xi$. The derivative is almost constant with $\xi$, indicating that the exponent $\gamma$ in the scale-free asymptotic correlation is very close to zero. **B.** Same as in **A** for the speed correlation.

Figure 4

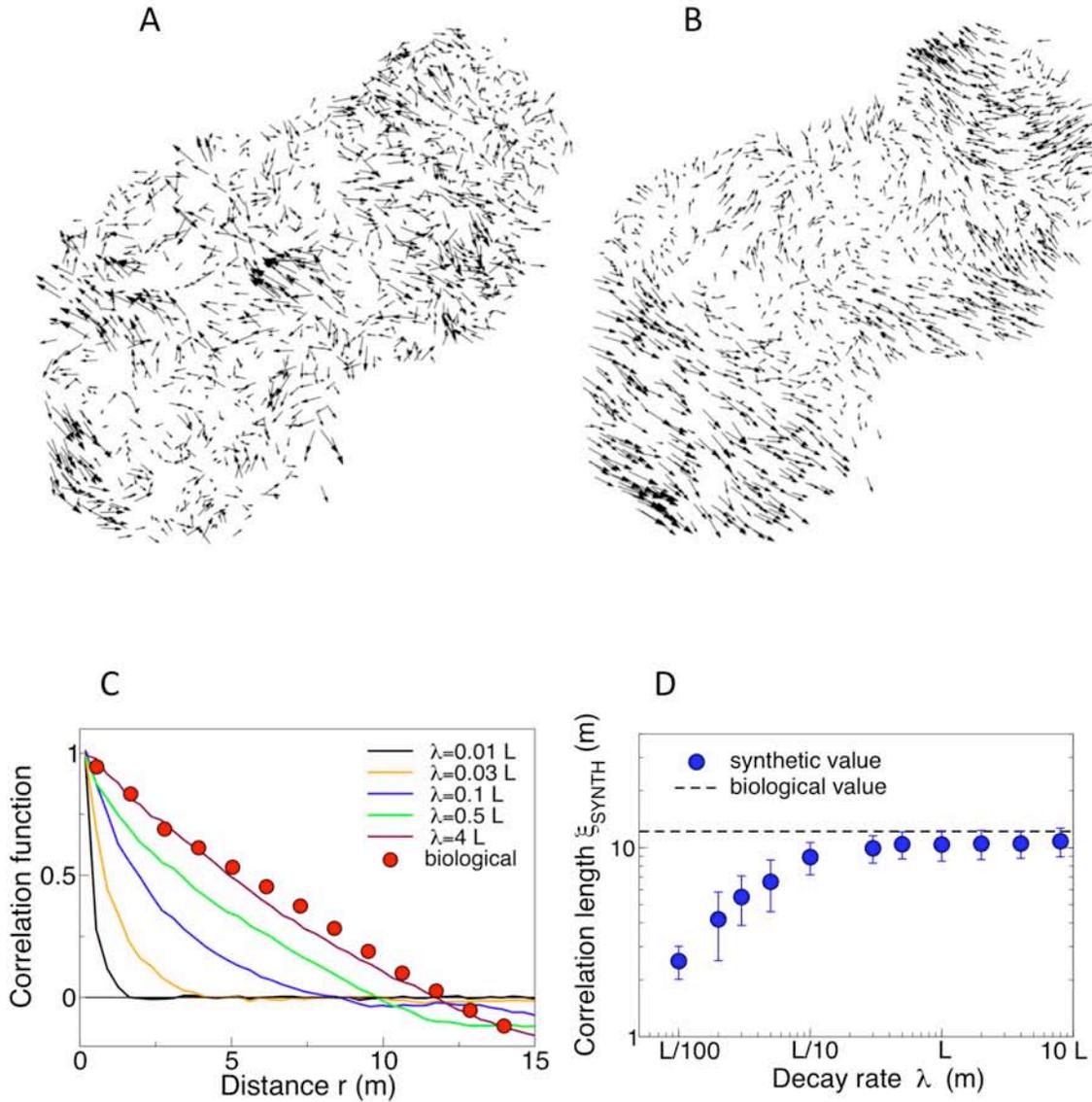

**Fig.4:** Random synthetic velocities. In each flock we replace the actual birds' velocity fluctuations with a set of synthetic random vectors correlated over a length $\lambda$ that we can arbitrarily tune (see text). The synthetic fluctuations are located at the same positions as birds in a real flock (in this figure we used flock 28-10, the same as in Fig.1). **A**. Synthetic fluctuations in the non-scale-free case, $\lambda=0.05L$. The domains are quite small and have a size comparable to $\lambda$. **B.** Synthetic fluctuations for $\lambda=4L$. In this scale-free limit the domains are very similar to the actual biological ones displayed in Fig.1B. **C**. Synthetic correlation functions $\hat{C}(r,L)$ for various values of the decay length $\lambda$. By increasing $\lambda$ the synthetic correlation function becomes more and more long-ranged and it finally becomes very close to the actual biological one in the scale-free regime $\lambda>L$. **D**. Synthetic correlation length $\xi_{SYNTH}$, as a function of the decay length $\lambda$ (each point is an average over 50 synthetic samples; errors bars are standard deviations). As long as $\lambda$ is smaller than the size of the flock $L$, $\xi_{SYNTH}$ grows following $\lambda$. However, in the scale-free regime, $\lambda>L$, $\xi_{SYNTH}$ saturates to a value very close to the actual biological one.

# SUPPORTING INFORMATION

## What is information transfer?

Our statement that scale-free correlation is necessary to transmit undamped information across the system seems to contradict the ordinary physical fact that *waves* can coherently travel for a length that is generally much larger than the correlation length of the system. Sound propagation in the air (where the correlation length is very short) is the most obvious example of such phenomenon. One may thus object that orientation or density waves propagating across the flock could transport information even in absence of scale-free correlations.

It is certainly true that a *perturbation* can propagate across the flock even with a short correlation length. However, this does not automatically mean that *information* is transferred. In order to transform perturbation into information a nontrivial process of encoding and decoding is needed: for example, information may be encoded in speech, which is propagated as sound, and decoded again into information by the recipient. If all is well, the state of the recipient is then changed in a useful and permanent way as a result of such information transfer. One may suggest that, similarly, a panic wave triggered by a predator at one end of the flock may travel to birds at the other end and change their cognitive state in a useful way ('switch to panic'). However, even though some information is actually transferred ('panic'), this is absolutely insufficient to make the flock respond collectively: the cognitive change of state of the birds is not telling them in what direction to turn, nor how to keep cohesion in the group, nor how to respond coherently. Travelling waves do not encode all this essential information.

A far more effective way to transmit information consists in transferring a perturbation that changes the state of the individuals in a *directly* useful and *permanent* way, without the need of any process of decoding: for example, a bird changes direction of motion as a result of an attack, heading away from the predator. This change is transferred to all correlated birds, so that they all change their heading in a similar and permanent way. There is no need to interpret the transferred perturbation, as it is directly useful to each recipient to get away from the predator. In this case perturbation and information are the same thing. Unlike a passing wave, which in absence of decoding leaves unaltered the final state of the bird, information transferred through correlation makes a permanent change in the dynamical state of the individual, which is essential to achieve collective response.

In physical terms, we may say that, in absence of decoding, what is needed to transfer permanent information is a finite response at zero frequency, namely a finite static susceptibility, the span of which is regulated by the static correlation length. For example, oscillatory speed waves (nonzero frequency) could propagate through the flock, but they would leave unaltered the dynamical state of the birds after they have passed. In order to change the speed of the entire flock *permanently* (and thus usefully) one needs a finite response at zero frequency of the speed fluctuations, and hence a scale-free correlation length of the modulus. This is exactly what we found in starling flocks.

We therefore conclude that the only way to transfer directly useful information in a permanent way to *all* individuals in the flock, with no need of a complex neural process of encoding/decoding, is indeed to have scale-free correlation of the entire dynamical states of the birds.

**Spontaneous symmetry breaking and Goldstone's theorem**

As we have seen the polarization of a flock is nonzero (it is in fact very large), and this means that velocities are globally ordered. From a physicist's perspective this can be rephrased by saying that there is spontaneous breaking of the rotational symmetry. It may then be objected that the presence of scale-free correlations is a mere consequence of this symmetry breaking. Indeed, in physics whenever a continuous symmetry (as the rotational one) is spontaneously broken, giving rise to a nonzero order parameter (as the global velocity), it is possible to prove that the fluctuations transverse to the direction of the order parameter are scale-free correlated. This is Goldstone's theorem [1]. For example, a magnet below the critical temperature develops a nonzero global magnetization; as a consequence, fluctuations of the spins' orientations have infinite correlation length [2]. For this same reason, one may also object that long-range correlation is not, by itself, a symptom of criticality, because when a continuous symmetry is spontaneously broken there are scale-free correlations also below the critical point.

There are two replies to these objections. First, Goldstone's theorem can only be proven for Hamiltonian systems. It is far from obvious whether it holds or not for a flock of birds, a highly complex system of which we ignore about everything. Second, when a continuous symmetry is spontaneously broken, not *all* fluctuations are scale-free correlated. Coming to our case, fluctuations of the *modulus* of the velocity (speed) do not need to develop scale-free correlations according to Goldstone's theorem, because they have nothing to do with the rotational broken symmetry. However, as we have shown, we *do* find scale-free correlations also of the speed fluctuations. There is no obvious way to explain such correlation by using symmetry arguments, as Goldstone's theorem. This result is quite important, as it shows that in flocks all dynamical modes are scale-free correlated, not only those connected to the rotational broken symmetry. It really seems that flocks are critical in some fundamental way.

**Comparison with theoretical models**

The fact that correlations in flocks are scale-free seems to be consistent with some theoretical and numerical studies performed in self-propelled particle systems, where the correlation function was found to be a power law [3,4]. The comparison, however, is not straightforward. First, the theoretical studies of [3,4] were performed in presence of periodic boundary conditions, which is clearly not the experimental situation. Second, in these studies there were no bounded flocks, as 'birds' were sparse across the entire periodic system. Hence, all the features we have found for the correlation function, which are crucially related to the presence of flocks of various finite sizes, cannot be compared with the theoretical predictions. Finally, in most models the modulus of the velocity is a constant and therefore by construction there cannot be any correlation of the modulus fluctuations, in contrast with what we found here.

## Table S1. Polarization and other features in the analyzed flocks

Flocking events are labelled according to session number and to the position within the session they belong to. The polarization is defined in eq.(1) of the main text. The linear size *L* of the flock is defined as the maximum distance between two birds belonging to the flock. The velocity is that of the centre of mass, i.e. the mean velocity of the group. All values are averaged over several instants of time during the event.

| Event | Number of birds | Polarization | Velocity (m/s) | Flock's size L (m) |
|---|---|---|---|---|
| 16-05 | 2941 | 0.962 | 15.2 | 79.2 |
| 17-06 | 552 | 0.935 | 9.4 | 51.8 |
| 21-06 | 717 | 0.973 | 11.8 | 32.1 |
| 25-08 | 1571 | 0.962 | 12.1 | 59.8 |
| 25-10 | 1047 | 0.991 | 12.5 | 33.5 |
| 25-11 | 1176 | 0.959 | 10.2 | 43.3 |
| 28-10 | 1246 | 0.982 | 11.1 | 36.5 |
| 29-03 | 440 | 0.963 | 10.4 | 37.1 |
| 31-01 | 2126 | 0.844 | 6.8 | 76.8 |
| 32-06 | 809 | 0.981 | 9.8 | 22.2 |
| 42-03 | 431 | 0.979 | 10.4 | 29.9 |
| 48-17 | 871 | 0.886 | 11.2 | 31.1 |
| 49-05 | 797 | 0.995 | 13.9 | 19.2 |
| 54-08 | 4268 | 0.966 | 19.1 | 78.7 |
| 57-03 | 3242 | 0.978 | 14.1 | 85.7 |
| 58-06 | 442 | 0.984 | 10.1 | 23.1 |
| 58-07 | 554 | 0.977 | 10.5 | 19.1 |
| 63-05 | 890 | 0.978 | 9.9 | 52.9 |
| 69-09 | 239 | 0.985 | 11.8 | 17.1 |
| 69-10 | 1129 | 0.987 | 11.9 | 47.3 |
| 69-13 | 1947 | 0.937 | 9.6 | 44.8 |
| 69-19 | 803 | 0.975 | 13.8 | 26.4 |
| 72-02 | 122 | 0.992 | 13.2 | 10.6 |
| 77-07 | 186 | 0.978 | 9.3 | 9.1 |

# SUPPORTING FIGURES

Figure S1

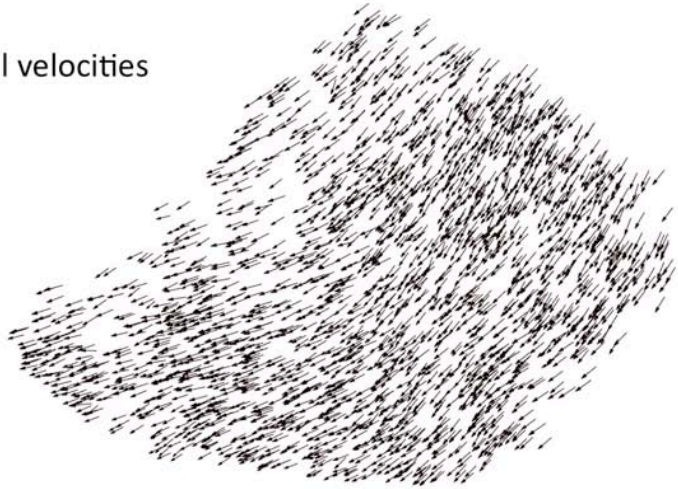

A: full velocities

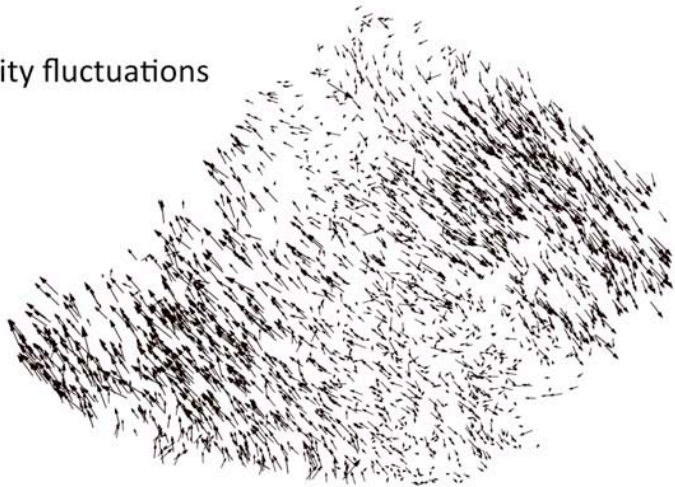

B: velocity fluctuations

**Figure S1 A.** This is the 2D projection of the velocities of the individual birds within a starling flock (event 57-03; 3242 birds - vectors scaled for clarity). **B.** This is the 2D projection of the individual velocity fluctuations in the same flock at the same instant of time as in A (vectors scaled for clarity). Large domains of strongly correlated birds are clearly visible. We recall that the modulus of the fluctuations is in fact much smaller than that of the full velocities (see Fig.1C in main text).

Figure S2

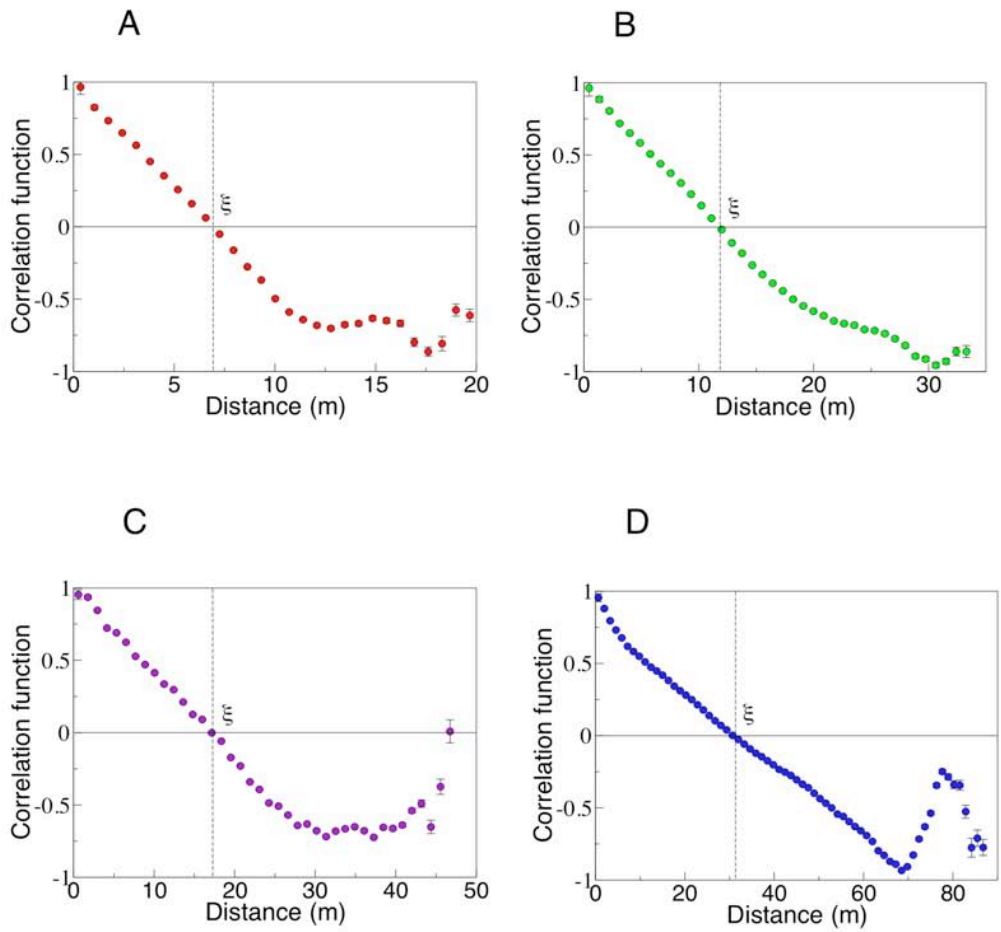

**Figure S2.** The correlation function C(r) in four other flocks: event 49-05 (panel **A**); event 25-10 (panel **B**); event 69-13 (panel **C**); event 57-03 (panel **D**). The behaviour of the correlation function is quite stable across different flocks; only for very large values of the distance $r$ boundary effects take over and the specific shape of the flock dominates the behaviour of the correlation function.

Figure S3

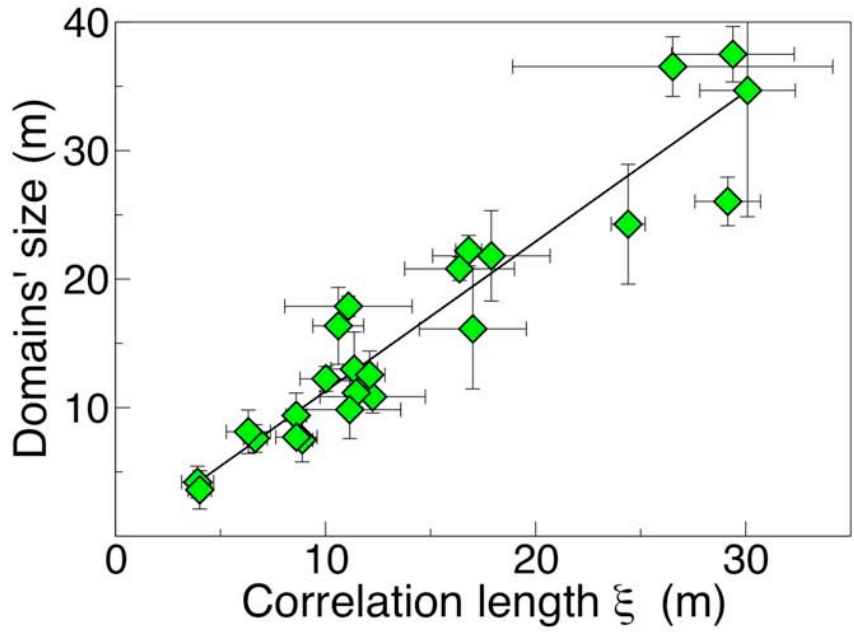

**Figure S3.** Size of the domains calculated form the covariance matrix vs. $\xi$, i.e. the zero of the correlation function, for all analyzed flocks. There is a clear linear correlation (Pearson correlation test: n=24, r=0.95, $P<10^{-11}$). The angular coefficient is equal to 1.16. This shows that the correlation length $\xi$ is indeed an excellent estimator of the domains' size.

## SUPPORTING REFERENCES

1. Goldstone J., Field theories with superconductor solutions. *Nuovo Cimento* **19**, 154 (1961).

2. Ryder, L.H. 1985 *Quantum Field Theory* (Cambridge University Press, Cambridge). 3. Tu, Y., Toner, J. & Ulm, M. Sound waves and the absence of Galilean invariance in flocks. *Phys. Rev. Lett.* **80**, 4819-4822 (1998).

4. Czirok, A. & Vicsek, T. Collective behavior of interacting self-propelled particles. *Physica A* **281**, 17-29 (2000).